\documentclass[
superscriptaddress,
reprint,
 amsmath,amssymb,
 aps,
 pra,
]{revtex4-2}

\usepackage{graphicx}% Include figure files
\usepackage{amsmath}
\usepackage{amsfonts}
\usepackage{amssymb}
\usepackage{bm}% bold math
\usepackage{dsfont}
\usepackage[dvipsnames]{xcolor}
\usepackage{siunitx}
\usepackage[%
]{hyperref}
\usepackage{mathtools}
\usepackage[normalem]{ulem}
\usepackage{inputenc}
\usepackage[T1]{fontenc}

\graphicspath{{./figures/}}

\definecolor{mmagenta}{cmyk}{0,1,0,0.12}

\usepackage{extdash}
\usepackage{csquotes}
\usepackage[caption=false,font=footnotesize]{subfig}

\usepackage{acro}
\acsetup{%
	first-style=long-short,
}

\DeclareAcronym{fdr}{%
	short = FDR,
	long = fluctuation\Endash dissipation relation,
	short-indefinite = an,
}

\DeclareAcronym{nisq}{%
	short = NISQ,
	long = noisy intermediate-scale quantum,
}

\DeclareAcronym{eth}{%
	short = ETH,
	long = eigenstate thermalization hypothesis,
 	short-indefinite = an,
}

\newcommand*{\diff}{\mathop{}\!\mathrm{d}}

\DeclarePairedDelimiterX{\commutator}[2]{[}{]}{#1,#2}
\DeclarePairedDelimiterX{\anticommutator}[2]{\{}{\}}{#1,#2}
\DeclarePairedDelimiter{\norm}{\lVert}{\rVert}

\DeclareMathOperator{\real}{Re}
\DeclareMathOperator{\imag}{Im}

\DeclarePairedDelimiter{\ket}{|}{\rangle}
\DeclarePairedDelimiter{\bra}{\langle}{|}
\DeclarePairedDelimiter{\braket}{\langle}{\rangle}
\DeclarePairedDelimiterX{\overlap}[2]{\langle}{\rangle}{#1\delimsize\vert\mathopen{}#2}
\DeclarePairedDelimiterX{\matrixelement}[3]{\langle}{\rangle}{#1\,\delimsize\vert\,\mathopen{}#2\,\delimsize\vert\,\mathopen{}#3}
\DeclareMathOperator{\variance}{Var}

\usepackage[capitalise]{cleveref}

\usepackage{mleftright}
\mleftright % fix spacing issues after \left \right commands
\usepackage{tikz}
\usetikzlibrary{quantikz}

\usepackage{xspace}

\newif\ifcomments
\commentstrue % comment out to hide comments

\definecolor{mypurple}{rgb}{0.49,0.18,0.56}
\definecolor{mygreen}{rgb}{0,0.5,0}
\definecolor{myblue}{rgb}{0,0,0.75}
\definecolor{mymagenta}{cmyk}{0,1,0,0.12}
\definecolor{mygray}{rgb}{0.5,0.5,0.5}

\begin{document}

% \title{Efficient measurement protocol for dynamical correlation functions in many-body systems on qudit devices} 
\title{Qudit-native measurement protocol for dynamical correlations using Hadamard tests}

\date{\today}

\author{Pavel P. Popov}
\email{pavel.popov@icfo.eu}
\affiliation{ICFO - Institut de Ciencies Fotoniques, The Barcelona Institute of Science and Technology, Av.\ Carl Friedrich Gauss 3, 08860 Castelldefels (Barcelona), Spain}

\author{Kevin T. Geier}
\affiliation{Pitaevskii BEC Center, CNR-INO and Dipartimento di Fisica, Universit\`a di Trento, 38123 Trento, Italy}
\affiliation{Trento Institute for Fundamental Physics and Applications, INFN, 38123 Trento, Italy}
\affiliation{Quantum Research Center, Technology Innovation Institute, P.O. Box 9639, Abu Dhabi, United Arab Emirates}

\author{Valentin Kasper}
\affiliation{Kipu Quantum GmbH, Greifswalderstr. 226, 10405 Berlin, Germany }

\author{Maciej Lewenstein}
\affiliation{ICFO - Institut de Ciencies Fotoniques, The Barcelona Institute of Science and Technology, Av.\ Carl Friedrich Gauss 3, 08860 Castelldefels (Barcelona), Spain}
\affiliation{ICREA, Pg.\ Lluis Companys 23, 08010 Barcelona, Spain}

\author{Philipp Hauke}
\affiliation{Pitaevskii BEC Center, CNR-INO and Dipartimento di Fisica, Universit\`a di Trento, 38123 Trento, Italy}
\affiliation{Trento Institute for Fundamental Physics and Applications, INFN, 38123 Trento, Italy}

\begin{abstract}
% Quantum processors based on qudits offer a hardware-efficient embedding of many complex quantum algorithms, but not all measurement schemes applicable to two-level systems naturally extend to their higher-dimensional counterparts.
% Quantum processors based on qudits offer hardware-efficient embeddings of many complex quantum algorithms.

Dynamical correlations reveal important out-of-equilibrium properties of the underlying quantum many-body system, yet they are notoriously difficult to measure in experiments. While measurement protocols for dynamical correlations based on Hadamard tests for qubit quantum devices exist, they do not straightforwardly extend to qudits.
Here, we propose a modified protocol to overcome this limitation by decomposing qudit observables into unitary operations that can be implemented and probed in a quantum circuit.
% Here, we propose a measurement protocol for dynamical correlations on qudit quantum hardware based on modified Hadamard tests, where qudit observables are implemented in the quantum circuit.
% By representing Hermitian operators as the real part of a unitary, we work out a scheme for the implementation of qudit observables and their dynamical correlations as quantum circuits.
We benchmark our algorithm numerically at the example of quench dynamics in a spin-1 XXZ chain with finite shot noise and demonstrate advantages in terms of signal-to-noise ratio over established protocols based on linear response.
% In a numerical simulation of quench dynamics in the spin-1 XXZ chain with finite shot noise, we compare our algorithm with protocols based on the linear response theory.
% We observe advantage of our protocol with respect to number of shots required for distinguishing the signal from noise.
% At last, comment on possible implementations of our protocol in existing qudit experiments.
Our scheme can readily be implemented on various platforms and offers a wide range of applications like variational quantum optimization and probing thermalization in many-body systems.

\end{abstract}

\maketitle

\section{Introduction}
The basic unit in which information is encoded in quantum computers is typically the qubit, the quantum generalisation of the classical bit.
However, many physical systems naturally offer units with a higher-dimensional local Hilbert space.
Embracing these higher dimensions leads to new ways for processing information based on qudits~\cite{Wang2020}, with applications like alternative quantum algorithms~\cite{Lanyon2009}, optimal measurements~\cite{Stricker2022}, and native encoding of complex quantum systems with higher local degrees of freedom~\cite{MacDonell2020}.
On a fundamental level, the different coherence~\cite{Ringbauer2018}, dissipation, and entanglement structure~\cite{Kraft2018} of qudit systems with respect to qubits can offer advantages for noise resilience~\cite{Cozzolino2019} and quantum error correction~\cite{Campbell2014}.
Thanks to these favorable properties, qudits have gained traction on various quantum technology platforms~\cite{Morvan2021,Chi2022,Ringbauer2022} with manifold applications, e.g., in quantum cryptography~\cite{Bruss1998,Thew2004} and quantum simulation~\cite{Hauke2012_rev,Preskill2018,Bharti2022}.
% Qudits have been recognized as useful for quantum cryptography~\cite{Bruss1998,Thew2004} and have gained traction in various quantum technology platforms~\cite{Morvan2021,Chi2022,Ringbauer2022}.
% This rapid progress allows for advanced quantum algorithms such as quantum simulation, making qudits valuable in the \ac{nisq} era~\cite{Hauke2012_rev,Preskill2018,Bharti2022}.
%
% \ckg{summarize advantages in one or two keywords, same for applications}

However, the new possibilities offered by qudit systems often come at the price of an increased complexity, such that not all quantum algorithms or measurement schemes generalize to qudits in a straightforward way.
This calls for the design of novel protocols tailored specifically for qudit systems.
Dynamical correlations involving observables at unequal times represent a class of observables that is inherently difficult to measure in quantum systems due to the collapse of the wave function.
These quantities play a fundamental role in statistical mechanics.
For example, they can be used to probe thermalization in isolated quantum systems via the so-called \ac{fdr}~\cite{Callen1951,Kubo1957,Kubo1966,Foini2011,Foini2012,Khatami2013,Orioli2019,Schuckert2020,Geier2022} and thus provide a test of the \ac{eth}~\cite{Deutsch1991,Srednicki1994,Nandkishore2015,Gogolin2016,DAlessio2016,Deutsch2018,Abanin2019,Moudgalya2022,Sierant2024}.
%
% Possible additional references to cite in the context of FDR as a probe of thermalization:
% 
% L. Foini, L. F. Cugliandolo, and A. Gambassi, Phys. Rev. B 84, 212404 (2011).
% L. Foini, L. F. Cugliandolo, and A. Gambassi, J. Stat. Mech. Theory Exp. 2012,
% P09011 (2012).
% E. Khatami, G. Pupillo, M. Srednicki, and M. Rigol, Phys. Rev. Lett. 111, 050403
% (2013).
% A. Piñeiro Orioli and J. Berges, Phys. Rev. Lett. 122, 150401 (2019).
% 
% Possible review to cite in the context of thermalization/ETH:
% R. Nandkishore and D. A. Huse, Annu. Rev. Condens. Matter Phys. 6, 15 (2015).
% C. Gogolin and J. Eisert, Reports on Progress in Physics 79, 056001 (2016).
% L. D’Alessio, Y. Kafri, A. Polkovnikov, and M. Rigol, Adv. Phys. 65, 239 (2016).
% J. M. Deutsch, Rep. Progr. Phys. 81, 082001 (2018).
% D. A. Abanin, E. Altman, I. Bloch, and M. Serbyn, Rev. Mod. Phys. 91, 021001 (2019).
% S. Moudgalya, B. A. Bernevig, and N. Regnault, Rep. Prog. Phys. 85 086501 (2022).
% 
Furthermore, in the context of variational quantum algorithms~\cite{Cerezo2021, Tilly2022}, many relevant observables like the gradient of the energy or the Fubini--Studi metric tensor~\cite{Wierichs2022} can be cast in the form of a dynamical correlation function~\cite{Li2017,McArdle2019,Yuan2019,Popov2024}.
% \ckg{list applications with citations}.
Various protocols exist for measuring dynamical correlations, but they are often efficient only for certain platforms, types of observables, and measurement precision~\cite{Knap2013,Yao2016,Uhrich2017,Elben2018,Vermersch2019,Schuckert2020,Geier2022}.
On digital qubit systems, the Hadamard-test~\cite{Ekert2002} has proven to be a versatile and robust method for accessing dynamical correlations~\cite{Bauer2016,Li2017,McArdle2019,Mitarai2019, Huggins2022, Libbi2022,Baroni2022, Mueller2023}.
However, these protocols strongly rely on the fact that Pauli operators on qubits are both Hermitian and unitary, which is not the case for qudits.

In this work, we formulate the Hadamard-test procedure for qudit systems on deterministic quantum circuits, enabling measurements of dynamical correlation functions in many-body systems with a higher-dimensional local Hilbert space.
% We recognize that a Hermitian matrix can be written as a sum of a unitary matrix and its Hermitian conjugate.
% Using this, we rewrite the dynamical correlation function of arbitrary observables as a sum of quantities that can be measured with Hadamard tests on the quantum device.
This is achieved by decomposing dynamical correlation functions of arbitrary observables into sums of unitaries, which can be measured with Hadamard tests on the quantum device, requiring only a single ancillary qubit.
% We compare its performance with protocols based on the theory of linear response and show an advantage in measuring the dynamical two-point commutator and anti-commutator in a spin-1 XXZ chain.
We numerically test the performance of our protocol by simulating a measurement of the two-time commutator and anti-commutator in a spin-1 XXZ chain with finite shot noise, where our scheme proves to be more efficient than established linear-response-based techniques~\cite{Kubo1957,Kubo1966}.
Our measurement protocol can readily be implemented on various qudit quantum devices, e.g., based on trapped ions, superconducting circuits, or neutral atoms, and thus provides a versatile tool for characterizing many-body physics on these platforms.

\begin{figure*}[t]
    \centering
\includegraphics[width=\linewidth]{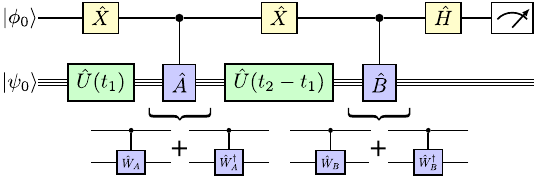}\\
    \caption{\textbf{Quantum circuit for measuring dynamical correlations based on Hadamard tests, extended to qudit devices.}
    The general structure consists of a qudit register that implements the physical system of interest plus an ancillary qubit. The register is initialized in state $\ket{\psi_0}$ and evolves under the unitary time-evolution operator $\hat{U}(t)$.
    The ancilla is subject to bit flip gates~$\hat{X}$ and the Hadamard gate~$\hat{H}$.
    In order to obtain the real and the imaginary parts of the dynamical correlation~$\braket{\hat{A}(t_1) \hat{B}(t_2)}$, the ancillary qubit is initialized in the state $\ket{\phi^{\pm}_0} = (\ket{0}+e^{i\alpha_{\pm}}\ket{1}) / \sqrt{2}$ with $\alpha_{+} = 0$ and $\alpha_{-} = \pi/2$, respectively.
    The decomposition of the qudit observables $\hat{A}$ and $\hat{B}$ into a sum of two unitaries $\hat{W}_X$ and $\hat{W}_X^\dagger$ with $X \in \{ A, B \}$, makes it possible to implement their action as controlled quantum gates between the ancilla and the register.
    After the circuit is executed, measuring the probability of finding the ancilla in the state $\ket{0}$ gives access to the desired value of the dynamical correlation function.% according to \cref{eq:probability_corr}.%
    }
    \label{fig:protocol}
\end{figure*}

 This paper is organized as follows.
In \cref{sec:measurement_protocol}, we present the general measurement protocol in a platform-agnostic framework.
\Cref{sec:numerical_results} puts forward our numerical results illustrating the Hadamard-test protocol for measuring dynamical correlations in a spin-1 XXZ chain with finite shot noise, followed by a comparison to linear-response-based schemes and a discussion of perspectives for an experimental realization.
Finally, we present our conclusions in \cref{sec:conclusion}.

\section{Measurement protocol}% based on Hadamard tests}
\label{sec:measurement_protocol}

% The protocol we propose in this work allows for measuring dynamical correlation functions in many-body systems with arbitrary local Hilbert space dimension. It is based on an implementation of deterministic quantum circuits on a system of a register that implements the physical system of interest and an ancillary qubit. The protocol can be realized in state-of-the-art qudit-based quantum devices like universal quantum computers and quantum gas microscope experiments. 

Our goal is to measure dynamical correlation functions of two Hermitian (multi\nobreakdash-)qudit observables $\hat{A}$ and $\hat{B}$ of the form
\begin{align}
    C_{AB}(t_1,t_2) = \braket[\big]{\hat{A}(t_1) \hat{B}(t_2)} \,,
\label{eq:dynamical_correlations}
\end{align}
% \ckg{$\braket{X Y}_c = \braket{XY} - \braket{X}\braket{Y}$ is a shorthand notation for the connected correlator (or cumulant)}
where we use the shorthand notation $\hat{O}(t) = \hat{U}^{\dagger}(t)\hat{O}\hat{U}(t)$.
The unitary operator~$\hat{U}(t)$ may equally well describe a Hamiltonian time evolution (e.g., in cold-atom experiments) or a sequence of quantum gates on a digital quantum device, for example, representing a parametrized quantum circuit in the context of variational quantum simulation algorithms~\cite{Li2017}.
% In the context of variational quantum simulation algorithms, instead of  a time evolution operator, we typically have a unitary operator that corresponds to (a part of) a parametrized quantum circuit~\cite{Li2017}.
%
In general, the two-time correlator in \cref{eq:dynamical_correlations} is a complex-valued quantity,
% and therefore only its real and imaginary part can be measured in the experiment.
$C_{AB} = C_{AB}^{+} / 2 - i C_{AB}^{-} / 2$, whose real and imaginary parts are given, respectively, by the two-time anti-commutator
\begin{subequations}
\begin{align}
    C_{AB}^{+}(t_1, t_2) = \braket[\big]{\anticommutator[\big]{\hat{A}(t_1)}{\hat{B}(t_2)}} \, \\
\intertext{and commutator}
    C_{AB}^{-}(t_1, t_2) = i\braket[\big]{\commutator[\big]{\hat{A}(t_1)}{\hat{B}(t_2)}} \,.
\end{align}
\label{eq:real_imag_correlator}
\end{subequations}
% Explicitly, we can split $C_{\hat{A}\hat{B}}$ into real and imaginary part 
% \begin{align}
%     C_{\hat{A}\hat{B}}(t_1,t_2) = \frac{1}{2}\braket{\anticommutator{\hat{A}(t_1)}{\hat{B}(t_2)}} - \frac{i}{2}i\braket{\commutator{\hat{A}(t_1)}{\hat{B}(t_2)}},
% \end{align}
% \ckg{do not use curly braces for visual appeal}
% given by the unequal-time anti-commutator~$C_{\hat{A}\hat{B}}^{+}$ \ckg{below, $C^+$ denotes the \emph{connected} anti-commutator} and commutator~$C_{\hat{A}\hat{B}}^{-}$, respectively.
% Our protocol allows one to measure $C^{\pm}_{AB}$ independently of each other by simply adjusting the initial state. 
% and unlike state-of-the-art deterministic protocols, with the same computational complexity. \ckg{move the last comment to comparison with linear response}

\Cref{fig:protocol} illustrates the quantum circuit implementing the measurement protocol for dynamical correlations.
The qudit register representing the physical system of interest is initialized in the state~$\ket{\psi_0}$.
Its evolution under the operator $\hat{U}(t)$ is interrupted at the times $t_1$ and $t_2$ by controlled gates coupling the register to an ancillary qubit.
% An ancillary qubit is coupled to the register through controlled gates.
These gates represent the action of observables $\hat{A}$ and $\hat{B}$, whose dynamical correlation we intend to measure.
The value of the corresponding correlation function is obtained by measuring the probability of the ancillary qubit to be in the computational $\ket{0}$ state at the end of the circuit.
So far, the protocol corresponds to the standard Hadamard-test procedure well-known for qubits~\cite{Li2017}, which strongly relies on the fact that observables are both Hermitian and unitary.
By contrast, for qudits, operators $\hat{A}$ and $\hat{B}$ are typically not unitary and therefore do not directly correspond to physical quantum gates.
% A closer look to the proposed circuit reveales that the controlled operations are in fact not physical - they do not correspond to controlled unitaries and thus quantum gates, since the operators $\hat{A}$ and $\hat{B}$ are Hermitian, but not unitary.

% Since the qudit observables $\hat{A}$ and $\hat{B}$ are not necessarily unitary, it is not straight-forward to implement their controlled versions as quantum gates.
In order to extend such a protocol to qudits, we represent $\hat{A}$ and $\hat{B}$ as a sum of unitary operations that can be inserted in the quantum circuit. The procedure can be seen as a special case of the more general linear combination of unitaries (LCU) method~\cite{Chakraborty2024}. % of $\hat{A}$ and $\hat{B}$.
In general, we can decompose a Hermitian qudit operator~$\hat{X}$ into a sum of a unitary operator~$\hat{W}_X$ and its Hermitian conjugate as
\begin{align}
    \hat{X} = \frac{1}{2}\norm[\big]{\hat{X}}(\hat{W}_X + \hat{W}_X^{\dagger}) \,,
\label{eq:unitary_decomposition}
\end{align}
% \begin{subequations}
% \label{eq:unitary_decomposition}
% \begin{align}
%     \hat{A} &= \frac{1}{2}\norm[\big]{\hat{A}}(\hat{W}_A + \hat{W}^{\dagger}_A),\\
%     \hat{B} &= \frac{1}{2}\norm[\big]{\hat{B}}(\hat{W}_B + \hat{W}^{\dagger}_B),
% \end{align}
% \end{subequations}
% \ckg{typeset norm as $\norm{\hat{X}}$}
where $\norm{\cdot}$ denotes the spectral norm of the operator (given by the eigenvalue with the largest magnitude).
% \ckg{Do we need to specify the norm precisely?}
The above relation can readily be inverted, yielding
% We define the unitary $\hat{W}_X$ for a general Hermitian operator $X$ as
% \begin{align}
%     \hat{W}_X = \frac{1}{2}(\hat{X}/\norm{\hat{X}}) + \frac{i}{2}\sqrt{\mathbb{1}- (\hat{X}/\norm{\hat{X}})^2}.
% \end{align}
% 
\begin{align}
    \hat{W}_X = \dfrac{\hat{X}}{\norm{\hat{X}}} + i\sqrt{\mathds{1}- \dfrac{\hat{X}^2}{\norm{\hat{X}}^2}} \,.
\label{eq:unitary_W}
\end{align}
% \ckg{use display style fractions}
% The construction of U can be done explicitly following the recipe:
% \begin{itemize}
%     \item Diagonalise $S \rightarrow D_S = V^{\dagger}SV.$
%     \item Define $||S|| := \underset{i}{\max} |\lambda_i|$, where $\lambda_i$ are the diagonal elements of $D_S$.
%     \item Define the normalised $\tilde{D}_S := \frac{1}{||S||}D_S$.
%     \item Define the unitary matrix $W_D := \tilde{D}_S + i\sqrt{\mathbb{1}- \tilde{D}^2_S}.$
%     \item Define $W_S := \frac{1}{2}||S||VW_DV^{\dagger}$.
% \end{itemize}
% \ckg{formulate as text}

This decomposition allows us to rewrite \cref{eq:dynamical_correlations} %of the operators $\hat{A}$ and $\hat{B}$
as a sum of correlation functions of unitaries,
% By inserting Eq.~\eqref{eq:unitary_decomposition} into Eq.~\eqref{eq:dynamical_correlations}, we obtain
\begin{align}
    % C_{\hat{A}\hat{B}}(t_1,t_2) = &\norm[\big]{\hat{A}} \norm[\big]{\hat{B}}\big( \langle \hat{W}_A(t_1)\hat{W}_B(t_2)\rangle  \notag\\&+\langle \hat{W}_A(t_1)\hat{W}^{\dagger}_B(t_2)\rangle
    % +\langle \hat{W}^{\dagger}_A(t_1)\hat{W}_B(t_2)\rangle \notag\\&+\langle \hat{W}^{\dagger}_A(t_1)\hat{W}^{\dagger}_B(t_2)\rangle \big),
    C_{A B}(t_1,t_2) = \frac{1}{4}\norm[\big]{\hat{A}} \norm[\big]{\hat{B}}
    % \sum_{\substack{\hat{V}_1 = \hat{W}_{A}, \hat{W}_A^\dagger \\ \hat{V}_2 = \hat{W}_{B}, \hat{W}_{B}^\dagger}} C_{V_1 V_2}(t_1, t_2) .
    % \sum_{i,j \in \{ 1, \dagger \}} C_{W_A^i W_B^j}(t_1, t_2) .
    \sum_{\mathclap{V, V^\prime \in \{W, W^\dagger\}}} C_{V_A V_B^\prime}(t_1, t_2) \,.
    % \sum_{i = 1, 2} C_{V_i V_{i + 2}}(t_1, t_2)
    % \sum_{i = 1}^2 \sum_{j = 3}^4 C_{V_i V_j}(t_1, t_2)
    % \sum_{i,j = 1}^2 C_{X_i Y_j}(t_1, t_2)
\label{eq:qudit_dynamical_correlations}
\end{align}
% with $\hat{V}_1 = \hat{W}_A$, $\hat{V}_2 = \hat{W}_A^\dagger$, $\hat{V}_3 = \hat{W}_B$, and $\hat{V}_4 = \hat{W}_B^\dagger$.
% with $\hat{X}_{1} = \hat{W}_A$, $\hat{X}_2 = \hat{W}_A^\dagger$, $\hat{Y}_1 = \hat{W}_B$, and $\hat{Y}_2 = \hat{W}_B^\dagger$.
% \ckg{
% The sum in \cref{eq:qudit_dynamical_correlations} has only two terms, but should be four!
% I propose the following notation (see commented out code for an alternative):
% \begin{align*}
%     % C_{A B}(t_1,t_2) = \frac{1}{4}\norm[\big]{\hat{A}} \norm[\big]{\hat{B}} \hspace{1ex}
%     % \sum_{\mathclap{\substack{V_A \in \{ W_A, W_A^\dagger \} \\ V_B \in \{ W_B, W_B^\dagger \}}}} \hspace{1ex} C_{V_A V_B}(t_1, t_2)
%     C_{A B}(t_1,t_2) = \frac{1}{4} \norm[\big]{\hat{A}} \norm[\big]{\hat{B}} \sum_{\mathclap{V, V^\prime \in \{W, W^\dagger\}}} C_{V_A V_B^\prime}(t_1, t_2)
% \end{align*}
% (Instead of the complex correlator $C_{A B}(t_1,t_2)$, we could also write the expression for the (anti\nobreakdash-)commutator $C_{A B}^{\pm}(t_1,t_2)$, which is the quantity we are interested in.)
% }

%
% By taking the real or imaginary part on both sides of Eq.~\eqref{eq:qudit_dynamical_correlations}, we formulate the physically relevant real and imaginary part of $C_{\hat{A}\hat{B}}(t_1,t_2)$ as a sum of quantities, each of which can be measured by a distinct realization of the circuit in \cref{fig:protocol}. The controlled gates correspond now to the unitaries $\hat{W}^{(\dagger)}_A$ and $\hat{W}^{(\dagger)}_B$.
The real and imaginary parts of the correlators appearing on the right-hand side of \cref{eq:qudit_dynamical_correlations} can now be accessed by distinct realizations of the circuit in \cref{fig:protocol}, where the controlled gates correspond to the unitaries $\hat{W}_A^{(\dagger)}$ and $\hat{W}_B^{(\dagger)}$.
Let $\ket{\Psi_{V_A V_B}^{\pm}(t_1, t_2)}$ be the total quantum state after execution of the circuit, given the initial ancilla state $\ket{\phi_0^{\pm}} = (\ket{0} + e^{i\alpha_{\pm}}\ket{1}) / \sqrt{2}$ with $\alpha_+ = 0$ and $\alpha_- = \pi / 2$ and intermittent controlled unitaries $\hat{V}_{A} \in \{ \hat{W}_A, \hat{W}^{\dagger}_A \}$ and $\hat{V}_{B} \in \{ \hat{W}_B, \hat{W}^{\dagger}_B \}$.
As shown in \cref{app:proof_prob}, the value of the two-time (anti\nobreakdash-)commutator can be extracted from the probability
% \begin{align}
$P_{V_A V_B}^{\pm} (t_1, t_2) = |\overlap{0}{\Psi_{V_A V_B}^{\pm}(t_1, t_2)}|^2$
% \end{align}
of measuring the ancillary qubit in the state $\ket{0}$ according to
\begin{align}
    C_{A B}^{\pm}(t_1,t_2) = \norm[\big]{\hat{A}} \norm[\big]{\hat{B}}
    \sum_{\mathclap{V, V^\prime \in \{W, W^\dagger\}}} \Big( P_{V_A V_B^\prime}^{\pm}(t_1, t_2) - \frac{1}{2} \Big) \,.
\label{eq:probability_corr}
\end{align}
% \ckg{
% Should be:
% \begin{align*}
%     \sum_{\mathclap{V \in \{W, W^\dagger\}}} C_{V_A V_B}^{\pm}(t_1, t_2) = \sum_{\mathclap{V \in \{W, W^\dagger\}}} \left[ 4 P_{V_A V_B^\dagger}^{\pm}(t_1, t_2) - 2 \right]
% \end{align*}
% (we have $V_A$ and $V_B$ on the left-hand side, but $V_A$ and $V_B^\dagger$ on the right-hand side)
% \newline
% I would prefer to give instead the full expression for the (anti\nobreakdash-)commutator of the observables $A$ and $B$:
% \begin{align*}
%     % C_{A B}^{\pm}(t_1,t_2) = \norm[\big]{\hat{A}} \norm[\big]{\hat{B}} \hspace{1ex}
%     % \sum_{\mathclap{\substack{V_A \in \{ W_A, W_A^\dagger \} \\ V_B \in \{ W_B, W_B^\dagger \}}}} \hspace{1ex} \Big( P_{V_A V_B}^{\pm}(t_1, t_2) - \frac{1}{2} \Big)
%     C_{A B}^{\pm}(t_1,t_2) = \norm[\big]{\hat{A}} \norm[\big]{\hat{B}}
%     \sum_{\mathclap{V, V^\prime \in \{W, W^\dagger\}}} \Big( P_{V_A V_B^\prime}^{\pm}(t_1, t_2) - \frac{1}{2} \Big)
% \end{align*}
% }
% for $V_1 \in \{W_A, W_A^\dagger\}$ and $V_2 \in \{W_B, W_B^\dagger\}$.
% A proof of this relation can be found in Appendix~\ref{app:proof_prob}.
%
Therefore, the same circuit allows one to measure either the real or the imaginary part of the dynamical correlation function simply by preparing the ancillary qubit in a different initial state.
% \ckg{Other remarkable features of the protocol could be mentioned here.}

% Remarkably, this protocol allows us to give an analytical formula for the variance of the correlation function:
% 
% A quantity of practical importance is the variance of the dynamical correlation function in \cref{eq:qudit_dynamical_correlations}. A protocol that minimizes this variance is efficient in terms of number of measurements that need to be performed on the quantum device to obtain $C^{\pm}_{A B}$ with sufficient precision.
% \ckg{Good sentence, but it provokes the question whether our protocol is optimal in that sense, which I think we do not want to discuss...}
% 
A quantity of practical importance is the variance of the estimator for the dynamical correlation function in \cref{eq:qudit_dynamical_correlations}, as it determines the number of measurements that need to be performed on the quantum device to obtain $C^{\pm}_{A B}$ with sufficient precision.
\sout{In \mbox{\cref{app:variance}}, we derive the result}
If the probabilities $P_{V_A V_B}^{\pm}$ are estimated by repeating the Hadamard test with controlled unitaries $\hat{V}_A$ and $\hat{V}_B$, respectively, $M_{V_A V_B}$ times, the two-time (anti\nobreakdash-)commutator can be estimated with variance
% \sout{\begin{align}
%     \variance[C^{\pm}_{A B}] =  \norm[\big]{\hat{A}}^2 \norm[\big]{\hat{B}}^2
%     \sum_{\btext{\mathclap{V,V^\prime  \in \{W, W^\dagger\}}}}
%     P^{\pm}_{V_A V_B^\btext{\prime}} \left( 1 - P^{\pm}_{V_A V_B^\btext{\prime}} \right) \,.
%     % \sum_{\substack{\hat{V}_1 = \hat{W}_{A}, \hat{W}_A^\dagger \\ \hat{V}_2 = \hat{W}_{B}, \hat{W}_{B}^\dagger}}
%     % \left| \overlap*{0}{\Psi_{\hat{V}_1 \hat{V}_2}^{\pm}(t_1, t_2)} \right|^2\notag\\&\times\bigg(1-\left| \overlap*{0}{\Psi_{\hat{V}_1 \hat{V}_2}^{\pm}(t_1, t_2)} \right|^2\bigg).
    
% \label{eq:var_formula}
% \end{align}}
%
\begin{align}
    \variance[C^{\pm}_{A B}] =  \norm[\big]{\hat{A}}^2 \norm[\big]{\hat{B}}^2
    \sum_{\mathclap{V,V^\prime  \in \{W, W^\dagger\}}}
    \frac{P^{\pm}_{V_A V_B^\prime} \big( 1 - P^{\pm}_{V_A V_B^\prime} \big)}{M_{V_A V_B^\prime}} \,.
    % \sum_{\substack{\hat{V}_1 = \hat{W}_{A}, \hat{W}_A^\dagger \\ \hat{V}_2 = \hat{W}_{B}, \hat{W}_{B}^\dagger}}
    % \left| \overlap*{0}{\Psi_{\hat{V}_1 \hat{V}_2}^{\pm}(t_1, t_2)} \right|^2\notag\\&\times\bigg(1-\left| \overlap*{0}{\Psi_{\hat{V}_1 \hat{V}_2}^{\pm}(t_1, t_2)} \right|^2\bigg).
\label{eq:var_formula}
\end{align}%
%
%
% \ckg{%
% Should be
% \begin{align*}
%     \variance[C^{\pm}_{A B}] = 4 \norm[\big]{\hat{A}}^2 \norm[\big]{\hat{B}}^2
%     \sum_{\mathclap{V, V^\prime \in \{W, W^\dagger\}}}
%     P^{\pm}_{V_A V_B^\prime} \left( 1 - P^{\pm}_{V_A V_B^\prime} \right)
% \end{align*}
% }
The derivation of this result is reported \cref{app:variance}.
Although the precise value of the variance ultimately depends on the details of the system, a useful \textit{a priori} upper bound is given by the product of the square norms of observables $\hat{A}$ and $\hat{B}$,
$\variance[C^{\pm}_{A B}] \leq M^{-1} \norm{\hat{A}}^2 \norm{\hat{B}}^2$, assuming $M$ shots are allocated for measuring each configuration.
This result is consistent with the bound on the variance directly derivable from \cref{eq:real_imag_correlator} for a (hypothetical) projective measurement of the (anti\nobreakdash-)commutator with a total budget of $4 M$ shots, $\variance[C^{\pm}_{A B}] \leq (4M)^{-1} 4 \norm{\hat{A}}^2 \norm{\hat{B}}^2$.
In what follows, we illustrate our protocol for measuring dynamical correlations using numerical simulations and benchmark its performance with finite measurement statistics.
% We turn to the investigation of the performance of our protocol for measuring dynamical correlations in a numerical simulation including noise.

\section{Numerical results}
\label{sec:numerical_results}

\begin{figure*}[t!]
    \centering
    \subfloat{\label{fig:corr_fct_N10_lam02:a}\includegraphics[width=0.5\textwidth]{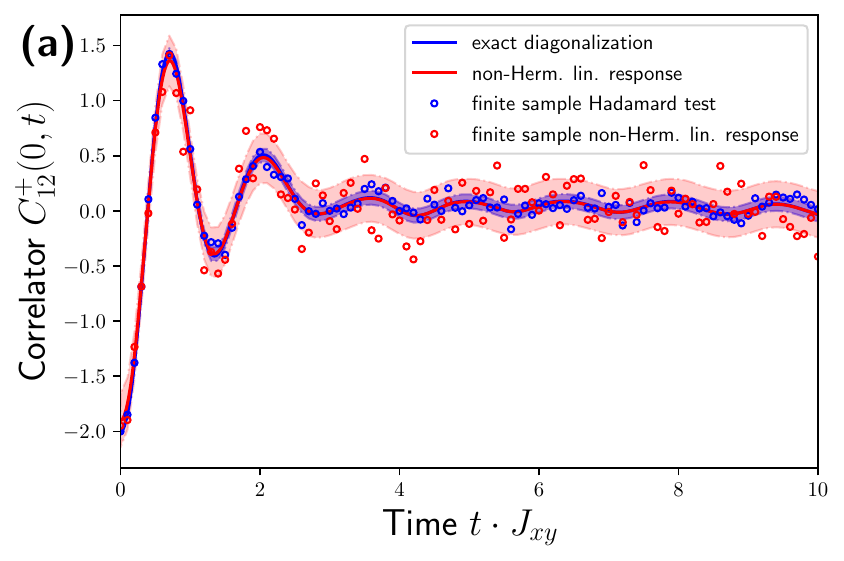}}%
    \subfloat{\label{fig:corr_fct_N10_lam02:b}\includegraphics[width=0.5\textwidth]{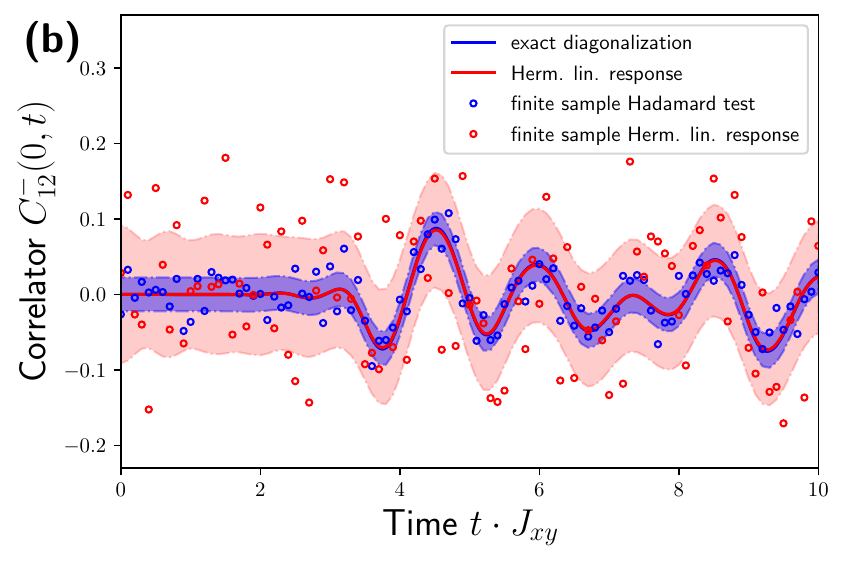}}%
    \caption{\textbf{
    Dynamical correlation functions after a quench in the spin-1 XXZ chain model.} The system of $N=10$ spins is initialized in the equal superposition of the two degenerate ground states of the model for $J_z / J_{xy} \to \infty$ and, after the quench, undergoes unitary evolution under the Hamiltonian~$H_{0}$ with $J_z / J_{xy} = 0.5$.
    The solid lines show the time trace of the two-time (a) anti-commutator and (b) commutator for the Hadamard-test protocol (blue) and for the linear-response protocol (red) with perturbation strength $\lambda = 0.2$ and a pulse area $J_{xy} \Delta t = 10^{-3}$.
    % The colored circles correspond to the expectation values obtained from sampling with finite statistics, illustrating the effect of shot noise for 1500 (blue and red) shots per point in case of the anti-commutator and 8000 (blue) and 12000 (red) shots per point for the commutator.
    The shaded areas mark the standard error of the mean computed directly from the quantum state, while the colored circles correspond to the expectation values obtained from sampling with finite statistics, illustrating the effect of shot noise.
    The Hadamard-test measurement of the anti-commutator (commutator) was sampled with 1500 (8000) shots per point, whereas the sampled linear-response measurement corresponds to 1500 (12000) shots per point.
    For a comparable total budget of shots, the Hadamard-test protocol achieves a better signal-to-noise ratio than the linear-response protocol throughout the simulation.%
    }
    \label{fig:corr_fct_N10_lam02}
\end{figure*}

In this section, we illustrate our protocol via numerical simulations for the example of quench dynamics in a spin-1 XXZ chain. This model represents a natural extension to qudit systems from its spin-$1/2$ version, which has served as a paradigmatic platform for testing novel quantum algorithms on qubit-based quantum hardware in the context of many-body quantum simulations~\cite{RevModPhys.93.025001}. Notably, closely related spin-1 models have been implemented on recently developed qudit quantum processors~\cite{edmunds2024}, representing promising targets for our proposed protocol. Furthermore, the spin-1 XXZ chain exhibits a richer phase diagram than its spin-$1/2$ counterpart, including the emergence of topological phases~\cite{Haldane1983a,Haldane1983b,Chen2003}.
These properties make the model appealing for demonstrating and validating our qudit-native approach for measuring dynamical correlations.

Apart from statistical errors, our protocol is exact in the sense that it does not involve approximations that could lead to systematic errors.
This is in contrast to protocols based on linear response, where by design a trade-off between accuracy and signal-to-noise ratio has to be made.
We therefore benchmark our technique for measuring dynamical correlations based on Hadamard tests with a particular focus on its performance in a finite-statistics setting compared to linear-response measurements.

% Our protocol is exact in the sense that it does not involve an approximation that has to be made in order to obtain the value of the dynamical correlation functions. However, it is not clear \ckg{sounds negative} how this protocol performs in a finite statistics setting, compared to other protocol for extracting dynamical correlation functions from measurements. Therefore, this section is dedicated to the application of our protocol and testing it in a realistic setup, taking into account finite statistics due to measurements. We compare our protocol with previous (state-of-the-art) deterministic techniques for the extraction of dynamical (anti-)commutators.

\subsection{Quench dynamics in a spin-1 XXZ chain}

% We employ a numerical simulation of quench dynamics in the spin-1 XXZ chain.
We consider an XXZ chain of $N$ spin-1 particles, whose Hamiltonian is given by
\begin{align}
\label{eq:xxz-hamiltonian}
    \hat{H} = \sum_{i=1}^{N-1} \left[ J_{xy} \left( \hat{S}^x_i \hat{S}^x_{i+1} + \hat{S}^y_i \hat{S}^y_{i+1} \right) + J_{z} \hat{S}^z_i \hat{S}^z_{i+1} \right] \,.
\end{align}
Here, $\hat{S}_i^\alpha$, with $\alpha \in \{ x, y, z \}$, are spin-1 matrices acting on spin~$i$, while $J_{xy}$ and $J_{z}$ denote the interaction strengths in the $xy$-plane and in $z$-direction, respectively.
% As shown in Ref.~\cite{Chen2003}, there are four different phases along the line of values for $J_z$.
% In the region of large $J_z / J_{xy}$, the ground state of the system is a (spin-1) Néel state, whereas for a small positive $J_z / J_{xy}$, the system is in the Haldane phase, which is characterized by zero-energy edge states~\cite{Jazdzewska2023}\cpp{Not sure what this means}.
% \ckg{Maybe better not to mention the Haldane phase in the first place.}
In order to initiate quench dynamics, we prepare the system deeply in the antiferromagnetic Néel phase, corresponding to the ground state for $J_z / J_{xy} \to \infty$. Since the ground state is degenerate in this limit, we choose a $\mathbb{Z}_2$-symmetric superposition of the two Néel states, but this choice is not crucial for our purposes.
The system is then evolved under the Hamiltonian~\labelcref{eq:xxz-hamiltonian} with $J_z / J_{xy} = 0.5$, which on the phase diagram lies in the bulk of the so-called Haldane phase~\cite{Chen2003}. While the initial state is characterized by antiferromagnetic order, which exhibits non-decaying spin--spin correlations, the Haldane phase is a gapped topological phase, characterized by non-vanishing string correlations. Moreover, it is considered a disordered phase; i.e., spin--spin correlations decay exponentially. Therefore, quenching from the antiferromagnetic phase to the Haldane phase, we expect to observe non-trivial dynamics in the two-point dynamical spin--spin correlations.

Specifically, we are interested in extracting the dynamical correlation functions in \cref{eq:real_imag_correlator} for the $\hat{S}^z$~operator of two distinct spins $i$ and $j$ in the chain, denoted by $C^{\pm}_{i j}(t_1, t_2) \equiv C_{S_i^z S_j^z}^{\pm}(t_1, t_2)$.
Following standard conventions~\cite{Berges2015}, we consider only the \enquote{connected part} of the two-time anti-commutator, $C^{+}_{i j}(t_1,t_2) \to C^{+}_{i j}(t_1,t_2) - 2 \braket{\hat{S}_i^z(t_1)} \braket{\hat{S}_j^z(t_2)}$ (the product of expectation values forming the \enquote{disconnected part} can be obtained from standard projective measurements).
% Specifically for the dynamical anticommutator, we subtract the product of expectation values of the time-evolved $S^z$-operators:
% \begin{align}
%     C^{+}_{zz}(t_1,t_2) = \langle \{S^z_i(t_1),S^z_j(t_2)\}\rangle - 2\langle S^z_i(t_1)\rangle\langle S^z_j(t_2)\rangle.
% \label{eq:stat_corr}
% \end{align}
% While the first term on the right hand side of Eq.~\eqref{eq:stat_corr} can be measured by our protocol, the second term is a product of expectation values, each of which can be measured by projective measurements.
%
In \cref{fig:corr_fct_N10_lam02}, we show the time trace of the two-time (anti\nobreakdash-)commutator $C_{12}^\pm(0, t)$ after the quench.

\subsection{Hadamard-test protocol with finite measurement statistics}

In this section, we investigate how well our Hadamard-test protocol reproduces the above ideal results with finite statistics.

\subsubsection{Implementation of the controlled unitaries for spin observables}
\label{sec:numerical_results:implementation}

In order to implement our protocol, we need to calculate the explicit form of the unitary operators in the decomposition of \cref{eq:unitary_decomposition}.
While for a general qudit observable~$\hat{X}$ the structure of the resulting unitary $\hat{W}_X$ may be complicated, the situation simplifies for spin observables.
The relevant operator for our case study is the single-spin $\hat{S}_z$ operator.
In the general spin-$S$ case, the matrix elements with respect to the eigenbasis of the $\hat{S}_z$~operator take the simple form
\begin{equation}
    \matrixelement[\big]{m}{\hat{W}_{S_z}}{m^\prime} = \delta_{m,m^\prime} \left( \frac{m}{S} + i \sqrt{1 - \frac{m^2}{S^2}} \right) \,,
\end{equation}
where $m, m^\prime \in \{ S, \dots, -S \}$. In particular, for $S = 1$,
we find in matrix representation
\begin{align}
    \hat{S}^z = \begin{pmatrix} 1 & 0 & 0\\ 0 & 0 & 0\\0 & 0 & -1 \end{pmatrix} \:\:\:\: \Rightarrow \:\:\:\: \hat{W}_{S_z} = \begin{pmatrix} 1 & 0 & 0\\ 0 & i & 0\\0 & 0 & -1 \end{pmatrix} \,.
\end{align}
The unitaries $\hat{W}_{S_x}$ and $\hat{W}_{S_y}$ can be obtained by applying the basis change that transforms $\hat{S}^{x}$ and $\hat{S}^{y}$ to $\hat{S}^z$, respectively.
We note that the controlled phase gate represented by $\hat{W}_z$ is generally available on state-of-the-art qudit quantum devices, in some cases even as a native operation~\cite{Meth2023}.
% \btext{It is important to highlight that the simplicity of the decomposition arises from the restriction to single-spin observables.}
Furthermore, for any many-body spin observable~$\hat{X}$, we can calculate $W_{\hat{X}}$ explicitly by using the representation of $\hat{X}$ as a sum of spin operator strings, as shown in \cref{app:n-body_obs}.

\subsubsection{Performance benchmark in the presence of shot noise}

In an experimental setting, the expectation values in Eq.~\eqref{eq:lin_resp} as well as the outcome of the Hadamard tests are subject to errors caused by finite measurement statistics.
We simulate these shot-noise errors by sampling the observables from the ideal distribution given by the total wave function of the system--ancilla register after the execution of the Hadamard-test circuit.
To this end, we assign a total budget of shots for measuring the four probabilities on the right-hand side of \cref{eq:probability_corr} as well as the two expectation values forming the disconnected part of the anti-commutator.

\Cref{fig:corr_fct_N10_lam02} compares the exact value of $C_{12}^\pm(0, t)$ (blue solid line) to the one obtained from sampling with finite statistics (blue circles).
% Already for a moderate number of $250$ ($2000$) shots for each expectation value, resulting in a total number of $1500$ ($8000$) shots per point, the shot noise is low enough to allow for an accurate measurement of the two-time anti-commutator (commutator).
The shot noise is low enough to allow for an accurate measurement of the two-time anti-commutator (commutator) already for a moderate number of $250$ ($2000$) shots for each expectation value, resulting in a total of $1500$ ($8000$) shots per point.

% The solid line correspond to the exact value of the corresponding quantity, whereas the empty circles correspond to the mean sampled value at each time step.
% The blue shaded area shows the standard error of the mean, calculated using the formula for the variance of a single measurement in Eq.~\eqref{eq:var_formula}.
% \ckg{such descriptions go in the caption}

% We have assigned a moderate amount of $n_s = 250$ shots per point for each quantity in the expression for the connected part of the two-time anti-commutator, resulting in $N_s = 1,5\times10^3 $ shots per point in total for the anti-commutator and $n_s = 2\times 10^3$ for each quantity of the commutator, resulting in  $N_s = 8\times10^3 $ shots per point in total for the commutator. The shot noise is low enough already at that number of shots and allows one to extract the signal of both dynamical correlation functions.

\subsection{Comparison to protocols based on linear response}

\begin{figure*}[t!]
    \centering
    \subfloat{\label{fig:lambda_plot:a}\includegraphics[width=0.5\textwidth]{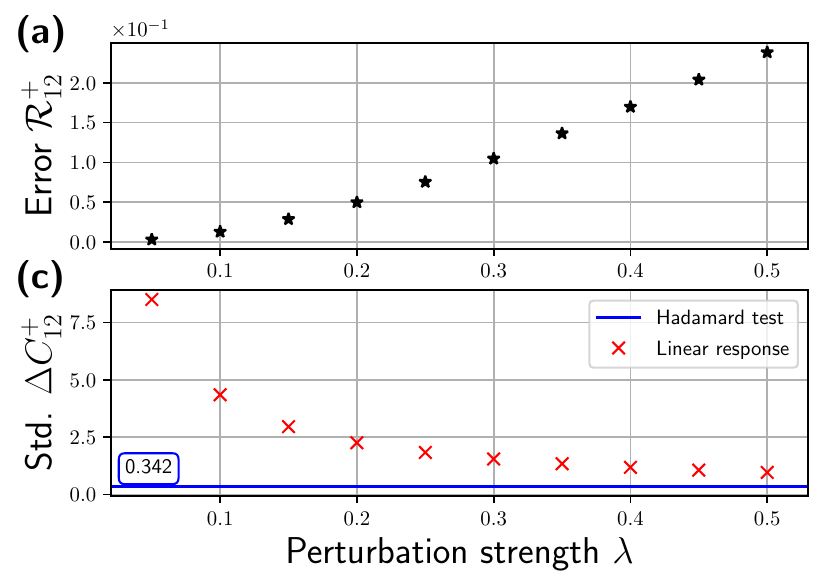}}%
    \subfloat{\label{fig:lambda_plot:b}\includegraphics[width=0.5\textwidth]{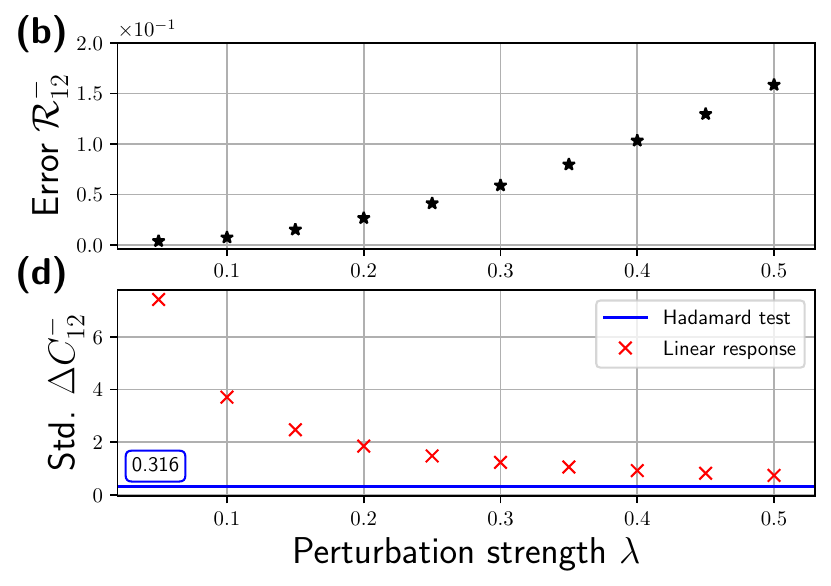}}%
    \caption{\textbf{
    Bias--variance tradeoff in the linear-response measurement of dynamical correlations.}
    With increasing perturbation strength~$\lambda$, the linear-response protocol incurs systematic errors in the two-time (a) anti-commutator and (b) commutator due to nonlinear effects.
    Concomitantly, the statistical error (standard deviation of a single measurement) of the (c) anti-commutator and (d) commutator decreases, requiring a trade-off between accuracy and signal-to-noise ratio.
    By contrast, the Hadamard-test protocol exhibits consistently lower statistical errors and does not introduce intrinsic systematic sources of error.}
    \label{fig:lambda_plot}
\end{figure*}

In this section, we compare our Hadamard-test protocol with other protocols for measuring dynamical correlation functions.
Many such protocols are specific to certain platforms or observables. 
To keep our analysis general, we focus the following discussion on protocols based on linear response theory~\cite{Kubo1957,Kubo1966}.
In fact, a large class of well-established protocols for measuring dynamical correlations, including non-invasive (or weak) measurements~\cite{Uhrich2017}, can be phrased in an implementation-independent way using this framework~\cite{Geier2022}.
% 
%\ckg{
% we exclude non-deterministic protocols like random unitaries because ??? 
%@Philipp, @Valentin:
%Should we try to argue explicitly why we did not consider random unitaries?
%Here are some possible arguments:
%\begin{itemize}
%\item focus only on \emph{deterministic} protocols
%\item random unitaries require a larger number of measurements due to the non-deterministic nature of the protocol (quantum statistics plus random unitary statistics necessary)
%\item HT and LR work locally, but most random unitaries work globally and require measurements of the entire system (although there might also be RU protocols that work locally)
%\item random unitaries inefficient for general qudit observables?
%\end{itemize}
%What do you think?
%}

\subsubsection{(Non-)Hermitian linear-response protocol}

As well known in linear response theory, the response of the system to a weak (Hermitian) perturbation gives access to the two-time \emph{commutator} according to Kubo's formula~\cite{Kubo1957,Kubo1966}.
More recently, measuring the two-time \emph{anti-commutator} by probing the linear response to a \emph{non-Hermitian} perturbation was proposed~\cite{Pan2020,Sticlet2022,Geier2022}.
Here, we employ such linear-response protocols to extract the response function directly in the time domain and benchmark the result against our Hadamard-test protocol.

Specifically, to measure the two-time (anti\nobreakdash-)commutator $C_{ij}^{\pm}(t_1, t_2)$, the system is first evolved under the unperturbed Hamiltonian~$\smash{\hat{H}_0}$ to the time $t_1$.
Then, a Hermitian (anti-Hermitian) perturbation of strength~$0 < \lambda \ll 1$ by the operator~$\hat{S}_j^z$ is applied in the form of a short rectangular pulse of duration~$\Delta t$, during which the system evolves under the perturbed Hamiltonian~$\hat{H}_\lambda = \hat{H}_0 - \lambda \hbar J_{xy} \hat{S}_j^z$ (in the non-Hermitian case, we have $\lambda \to i \lambda$).
After the perturbation is released, the system is evolved under the unperturbed Hamiltonian~$\smash{\hat{H}_0}$ up to time~$t_2$, and the expectation value of the observable $\hat{S}_i^z$ is measured, yielding the two-time (anti\nobreakdash-)commutator via
\begin{align}
    C^{\pm}_{ij, \, \mathrm{LR}}(t_1,t_2) &= \frac{1}{\lambda J_{xy} \Delta t} \big[ \braket[\big]{\hat{S}^z_i(t_2)}_{\pm} - \braket[\big]{\hat{S}^z_i(t_2)} \big] \,.
\label{eq:lin_resp}
\end{align}
Here $\braket{\cdots}_{\pm}$ and $\braket{\cdots}$ denote expectation values with respect to the perturbed and unperturbed states, respectively.
In the non-Hermitian case, we normalize the expectation value as $\braket{\cdots}_+ \to \braket{\cdots}_+ / \braket{\mathds{1}}_+$, accounting for the loss of probability to a complementary state space.

\subsubsection{Finite statistics and bias--variance trade-off}

As depicted in \cref{fig:corr_fct_N10_lam02}, the linear-response protocol accurately reproduces the time trace of the dynamical correlations for a sufficiently weak perturbation and infinite measurement statistics (red solid line). An advantage of our Hadamard-test protocol over the linear-response method becomes apparent if one takes statistical errors due to shot noise into account.
The red circles show the sampled values of the anti-commutator (commutator) using $750$ ($6000$) shots for each expectation value in \cref{eq:lin_resp}, corresponding to a total of $1500$ ($12000$) shots per point, which is comparable to the total budget used for the Hadamard-test protocol \footnote{In the non-Hermitian case, we reduce the number of shots according to the decrease of the norm of the state.}.
One can see that the Hadamard-test protocol yields a lower statistical variance and therefore a higher signal-to-noise ratio.

% For perturbation strength of $\lambda = 0.2$ and the duration of the perturbation $J_{xy} \Delta t = 10^{-3}$, the linear-response protocol gives accurate results, as depicted in~\cref{fig:corr_fct_N10_lam02} (for quantification, see next subsection). The solid red line corresponds to the ideal value of the dynamical (anti-)commutator, calculated by linear response theory, whereas the red circles - to the mean sampled value at each time step. The total number of shots per time step is the same as in the case of the Hadamard-test protocol, however, these shots are distributed to only two expectation values in accordance with Eq.~\eqref{eq:lin_resp}. The standard deviation of the mean value (red shaded area) is estimated from the statistical means from the sampled values.
% \ckg{I have some doubt whether the (non-)Hermitian LR protocol is introduced in sufficient detail such that the reader can appreciate the technical details in this paragraph.}

% Our next task is to systematically test some possible advantage of the Hadamard-test protocol in comparison to the linear-response protocol.
% As a metric, we use both the relative error of the linear-response protocol, defined as
In order to benchmark the relative performance of the Hadamard-test protocol and the linear-response protocol, we investigate the two figures of merit shown in \cref{fig:lambda_plot}:
the relative error
\begin{align}
    \mathcal{R}_{i j}^{\pm} = \frac{\int_0^{t}\diff t' \, |C^{\pm}_{ij}(0,t')-C^{\pm}_{ij, \, \text{LR}}(0,t')|^2}{\int_0^{t}\diff t^{\prime} \, |C^{\pm}_{ij}(0,t')|^2} \,,
\end{align}
% where the subscript \enquote{LR} stands for the correlation function calculated via Eq.~\eqref{eq:lin_resp}  and \enquote{exact} for the analytically calculated one,
measuring the systematic error of the linear-response protocol due to non-linear effects, and the time average of the standard deviation,
\begin{align}
\Delta C_{ij}^{\pm} = \frac{1}{t} \int_0^t \diff t^\prime \Delta C_{i j}^\pm(0, t^\prime) \,,    
\end{align}
measuring the statistical errors due to shot noise.

The linear-response protocol generally requires a trade-off between accuracy and signal-to-noise ratio.
On the one hand, the perturbation should be sufficiently weak that the response is in the linear regime, where nonlinear contributions to the response are negligible.
On the other hand, it should be sufficiently strong that the signal is resolvable against the background noise.
As can be seen in \cref{fig:lambda_plot}, the error grows with increasing $\lambda$, while the variance shrinks and vice versa.
This behavior can be viewed as a manifestation of the bias--variance tradeo-ff, which is well-known in statistics and machine learning.
% The linear-response protocol gives an accurate result for the dynamical correlation function within the so-called linear regime - where the perturbation amplitude is small. Going away from this regime, the relative deviation of the measured dynamical correlation function from its true value grows even with zero shot noise assumed, see the upper panels in \cref{fig:lambda_plot}. On the other hand, the standard deviation with which we measure the dynamical correlation function grows by making the perturbation smaller and smaller, see lower panels in \cref{fig:lambda_plot}. This represents a trade-off - we need perturbation small enough for the accuracy of the protocol and large enough in order to keep the standard deviation small.
% Such a trade-off does not exist in the case of the measurement protocol presented in this paper.
Importantly, there is no such trade-off in the case of our Hadamard-test protocol.
Furthermore, in the investigated quench scenario, the Hadamard-test protocol outperforms the linear-response protocol both in terms of accuracy and signal-to-noise ratio.
% 
% \nkg{\sout{In fact, as we mentioned earlier, there exists an upper bound for the variance, with which we estimate the dynamical correlation function, given by the norms of the observables in that correlation function (see Eq.~\eqref{eq:var_formula}).}}

\subsubsection{Experimental resources}

A fair comparison between our Hadamard-test protocol and the linear-response protocol also requires a discussion of the experimental resources required for their implementation.
In what follows, we discuss the cases of measuring the unequal-time commutator and anti-commutator separately.

For the commutator, the linear-response protocol requires a standard Hermitian perturbation, which either may be single-body, as in the case of spin--spin dynamical correlations, or may involve few-body interactions.
Importantly, the protocol typically admits ancilla-free realizations, where the perturbation is exerted directly on the system, e.g., in the form of a rapid change in the couplings in the Hamiltonian or as a unitary gate in a quantum circuit.
% In any case, realization of this perturbation does not necessarily involve additional overhead in terms of ancillary degrees of freedom or entangling operations.
By contrast, the protocol based on Hadamard tests requires an ancillary qubit and two entangling operations between this qubit and the system. 
Thus, depending on the device and scenario at hand, a decision has to be made between additional errors from shot noise and bias (for the linear-response protocol; see \cref{fig:lambda_plot}) and potential errors from faulty gates (for the Hadamard-test protocol).  
One way of circumventing the need for a physical ancilla for the Hadamard test is to use qudits with the number of controllable levels equal to double the local Hilbert space dimension of the system we want to simulate (e.g., $d=6$ in the example above).

% As for the dynamical anti-commutator, a possible implementation of the non-Hermitian linear-response protocol~\cite{Geier2022} would involve an ancillary qubit. By performing an entangling operation between the ancilla and the system and then post-selecting on the state of the ancilla, one can effectively engineer a non-Hermitian perturbation in the system. Therefore, the protocol presented here require similar resources.
% 
In our benchmarks of the anti-commutator, we have considered an implementation-agnostic \enquote{ideal} non-Hermitian linear-response measurement, where a non-Hermitian perturbation directly acts on the system degrees of freedom.
However, physical realizations of this scenario typically incur additional experimental overhead, e.g., entangling the system with an ancilla and performing postselected measurements on the ancilla state~\cite{Geier2022}.
Therefore, engineering an (approximate) effective non-Hermitian Hamiltonian requires resources similar to those of the Hadamard-test protocol.

Finally, we turn our attention to potential near-term experimental realizations of our Hadamard-test protocol for dynamical correlations of general observables. For systems with a large Hilbert space, implementing a general multi-qudit observable as a sum of unitaries presents significant practical challenges. While the decomposition in \cref{eq:unitary_decomposition} applies to any observable, the resulting unitary may have a complicated structure (unlike for spin observables; see \cref{sec:numerical_results:implementation}).
In this case, the gate complexity may be reduced by resorting to approximate decompositions, e.g., based on the LCU procedure~\cite{Chakraborty2024}, where one can trade off lower circuit depth for a more accurate approximation.
While in typical applications the circuit depth is dominated by the number of gates required to implement the system's time evolution, the added gate complexity due to the (approximate) implementation of the controlled unitaries may be of concern on noisy intermediate-scale quantum (NISQ) devices: In the presence of circuit noise, the maximum circuit depth is restricted by the finite coherence time.
This may require lowering the approximation accuracy of the controlled unitaries, introducing systematic errors, and it may impose limits on the times $t_1$ and $t_2$ of the dynamical correlations that can reliably be measured using the Hadamard test protocol. In situations where this effect is identified as a limiting factor, the linear-response protocol may provide a viable alternative since the implementation of the perturbation is typically favorable in terms of gate complexity.

\section{Conclusion}
\label{sec:conclusion}

In this work, we proposed an algorithm for measuring dynamical correlation functions on qudit quantum hardware.
Key to the scheme is the decomposition of general (multi-)qudit observables into unitaries, which can be exploited to represent dynamical correlations as the sum of quantities accessible via Hadamard tests.
We numerically benchmarked the performance of this protocol with finite measurement statistics using the example of a spin-1 XXZ model after a quench.
% Furthermore, we tested the performance of our measurement protocol by simulating finite shot noise sampling of dynamical correlation functions numerically.
% In the numerical simulation, the measurement protocol was able to reproduce the dynamical commutator and anti-commutator of spin observables in the spin-1 XXZ model after a quench.
Notably, in this scenario the Hadamard-test protocol outperforms established protocols based on linear response theory in terms of both accuracy and the number of shots needed to distinguish the signal from noise.

The Hadamard-test protocol for dynamical correlation functions can be extended in a straightforward way to measuring higher-order temporal correlations of the form $\braket{A_1(t_1) \cdots A_n(t_n)}$.
Such a generalization of our protocol would require only the implementation of deeper circuits and a higher number of controlled operations, while the structure of the circuit and the usage of the decomposition of Hermitian operators into a sum of unitaries remain the same. 

Finally, we emphasize that our Hadamard-test protocol for measuring dynamical correlations in qudit quantum systems can be used as a subroutine in existing algorithms with wide-ranging applications, ranging from thermalization in many-body systems~\cite{Schuckert2020,Geier2022,Zhou2022,Mueller2023} to the implementation of variational optimization protocols for the simulation of equilibrium and non-equilibrium properties of many-body systems~\cite{Yuan2019,Tilly2022,Popov2024}.
% \ckg{cite a selection of specific papers here?}
Moreover, the protocol is designed such that it can be implemented on presently available quantum simulation platforms, including trapped ions, cold atoms, and superconducting qubits.

\section{Acknowledgements}
The ICFO group acknowledges support from:
ERC AdG NOQIA; MCIN/AEI [PGC2018-0910.13039/501100011033,  CEX2019-000910-S/10.13039/501100011033, Plan National FIDEUA PID2019-106901GB-I00, the Plan National STAMEENA PID2022-139099NB-I00 project funded by MCIN/AEI/10.13039/501100011033 and by the “European Union NextGenerationEU/PRTR" (PRTR-C17.I1), FPI]; QUANTERA MAQS PCI2019-111828-2);  QUANTERA DYNAMITE PCI2022-132919 (QuantERA II Programme co-funded by the European Union’s Horizon 2020 program under Grant Agreement No. 101017733), Ministry of Economic Affairs and Digital Transformation of the Spanish Government through the QUANTUM ENIA project call – Quantum Spain project, and the European Union through the Recovery, Transformation, and Resilience Plan – NextGenerationEU within the framework of the Digital Spain 2026 Agenda; Fundació Cellex; Fundació Mir-Puig; Generalitat de Catalunya (European Social Fund FEDER and CERCA program, AGAUR Grant No. 2021 SGR 01452, QuantumCAT \ U16-011424, co-funded by ERDF Operational Program of Catalonia 2014-2020); Barcelona Supercomputing Center MareNostrum (FI-2023-1-0013); EU Quantum Flagship (PASQuanS2.1, 101113690); EU Horizon 2020 FET-OPEN OPTOlogic (Grant No. 899794); the EU Horizon Europe Program (Grant Agreement No. 101080086 — NeQST), ICFO Internal “QuantumGaudi” project; the European Union’s Horizon 2020 program under Marie Sklodowska-Curie Grant Agreement No. 847648; and  “La Caixa” Junior Leaders fellowships, La Caixa” Foundation (ID No. 100010434): CF/BQ/PR23/11980043. 

This project received funding from the European Union’s Horizon Europe research and innovation programme under Grant Agreement No. 101080086 NeQST, the European Union under NextGenerationEU PRIN 2022 Prot. n. 2022ATM8FY (CUP: E53D23002240006), the Italian Ministry of University and Research (MUR) through the FARE grant for the project DAVNE (Grant No. R20PEX7Y3A), the National Recovery and Resilience Plan (NRRP), Mission 4 Component 2 Investment 1.4 - Call for tender No. 1031 of the Italian Ministry for University and Research funded by the European Union – NextGenerationEU (Project No. CN\_00000013), and Project DYNAMITE QUANTERA2\_00056 funded by the Ministry of University and Research through the ERANET COFUND QuantERA II – 2021 call and co-funded by the European Union (H2020, Grant Agreement No. 101017733).
This work benefited from Q@TN, the joint laboratory between the University of Trento, Fondazione Bruno Kessler (FBK), National Institute for Nuclear Physics (INFN), and National Research Council (CNR).
We acknowledge support from Provincia Autonoma di Trento.
P.P.P. also acknowledges support from the “Secretaria d’Universitats i Recerca del Departament de Recerca i Universitats de la Generalitat de Catalunya” under Grant No. 2024 FI-3 00390, as well as the European Social Fund Plus.

Funded by the European Union. Views and opinions expressed are however those of the author(s) only and do not necessarily reflect those of the European Union, the European Commission, or the Italian Ministry of University and Research. Neither the European Union nor the granting authority can be held responsible for them.

\section*{Data Availability}

The data set used in this paper is available on Zenodo~\cite{popov_2025_15089902}.

\appendix

\section{Outcome of the quantum circuit implementing the Hadamard-test protocol}
\label{app:proof_prob}

In this appendix, we derive \cref{eq:probability_corr} by calculating the state of the system after applying each quantum gate in~\cref{fig:protocol}.
The system--ancilla register is initialized in the state
\begin{align}
    \ket{\Psi^{\alpha}_0} = \frac{1}{\sqrt{2}}(\ket{0} + e^{i\alpha}\ket{1})\otimes\ket{\psi_0}
\end{align}
with $\alpha \in \{0,\frac{\pi}{2}\}$, depending on whether we want to measure the unequal-time anti-commutator or commutator.
After flipping the ancilla, evolving the state up to time $t_1$, and applying the first controlled operation $\hat{V}_A$, we obtain the intermediate state
\begin{multline}
    % \ket{\Psi_{V_A}^\alpha(t_1)} = \\
    \frac{1}{\sqrt{2}}e^{i\alpha}\ket{0}\otimes \hat{U}(t_1)\ket{\psi_0}  + \frac{1}{\sqrt{2}}\ket{1}\otimes \hat{V}_A \hat{U}(t_1)\ket{\psi_0} \,.
\end{multline}
Next, we apply the second ancilla flip, evolve the register up to time $t_2$, and apply the second controlled operation $\hat{V}_B$, yielding
\begin{equation}
\begin{split}
    % &\ket{\tilde{\Psi}_{V_A V_B}^{\alpha}(t_1, t_2)} = \\
    \frac{1}{\sqrt{2}}\ket{0} &\otimes \hat{U}(t_2-t_1) \hat{V}_A \hat{U}(t_1) \ket{\psi_0} \\
    + \frac{1}{\sqrt{2}} e^{i\alpha} \ket{1} &\otimes \hat{V}_B \hat{U}(t_2)\ket{\psi_0} \,.
\end{split}
\end{equation}
Finally, after performing the basis transformation on the ancilla via the Hadamard gate, the final state before the measurement reads
\begin{equation}
\begin{split}
    &\ket{\Psi_{V_A V_B}^{\alpha}(t_1, t_2)} \\ =
    &\ket{0}\otimes\left[\frac{1}{2} \hat{U}(t_2-t_1) \hat{V}_A \hat{U}(t_1) \ket{\psi_0}  + \frac{1}{2}e^{i\alpha} \hat{V}_B \hat{U}(t_2) \ket{\psi_0}\right] \\
    + &\ket{1} \otimes \left[ \frac{1}{2} \hat{U}(t_2-t_1) \hat{V}_AU(t_1) \ket{\psi_0} - \frac{1}{2} e^{i\alpha} \hat{V}_B \hat{U}(t_2) \ket{\psi_0} \right] \,.
\end{split}
\end{equation}
% \ckg{Replace $V_1$ and $V_2$ consistently with $V_A$ and $V_B$.}
The probability for the ancillary qubit to be in the computational state $\ket{0}$ is therefore given by
\begin{multline}
    2 \left| \overlap*{0}{\Psi_{V_A V_B}^{\alpha}(t_1, t_2)} \right|^2 - 1 \\
    = \frac{1}{2} e^{-i\alpha} \matrixelement[\big]{\psi_0}{\hat{U}^{\dagger}(t_2) \hat{V}^{\dagger}_B \hat{U}(t_2-t_1) \hat{V}_A \hat{U}(t_1)}{\psi_0} + \text{c.c.} \,,
\end{multline}
where $\text{c.c.}$ denotes the complex conjugate.
By setting $\alpha = \alpha_+ \coloneqq{} 0$ and $\alpha = \alpha_- \coloneqq{} \pi / 2$, we get the real and the imaginary parts of the dynamical correlation function of $\hat{V}_A$ and $\hat{V}^{\dagger}_B$, respectively:

\begin{subequations}
\label{eq:probability_explicit}
\begin{align}
    2 P_{V_A V_B}^{+}(t_1, t_2) - 1 &= \frac{1}{2} \braket[\big]{\hat{V}_A^\dagger(t_1) \hat{V}_B(t_2) + \hat{V}_B^\dagger(t_2) \hat{V}_A(t_2)} \nonumber \\
    &= \real \braket[\big]{\hat{V}_B^\dagger(t_2) \hat{V}_A(t_1)} \,, \\
    2 P_{V_A V_B}^{-}(t_1, t_2) - 1 &= \frac{1}{2} i\braket[\big]{\hat{V}_A^\dagger(t_1) \hat{V}_B(t_2) -\hat{V}_B^\dagger(t_2) \hat{V}_A(t_1)} \nonumber \\
    &= \imag \braket[\big]{\hat{V}_B^\dagger(t_2) \hat{V}_A(t_1)} \,.
\end{align}
\end{subequations}
By means of the decomposition in \cref{eq:qudit_dynamical_correlations}, we can express the two-time (anti\nobreakdash-)commutator as
\begin{align}
    C_{AB}^{\pm} &= \frac{1}{4} \norm{\hat{A}} \norm{\hat{B}} \sum_{\mathclap{V, V^\prime \in \{W, W^\dagger\}}} \braket[\big]{e^{i \alpha_{\pm}} \hat{V}_A \hat{V}_B^\prime + e^{-i \alpha_{\pm}} \hat{V}_B^\prime \hat{V}_A} \nonumber \\
    &= \frac{1}{4} \norm{\hat{A}} \norm{\hat{B}} \sum_{\mathclap{V, V^\prime \in \{W, W^\dagger\}}} \braket[\big]{e^{-i \alpha_{\pm}} \hat{V}_B^{\prime \, \dagger} \hat{V}_A + e^{i \alpha_{\pm}} \hat{V}_A^\dagger \hat{V}_B^\prime} \nonumber \\
    &= \frac{1}{2} \norm{\hat{A}} \norm{\hat{B}} \sum_{\mathclap{V, V^\prime \in \{W, W^\dagger\}}} \big( \real \braket{\hat{V}_B^{\prime \, \dagger} \hat{V}_A} \cos \alpha_{\pm} \nonumber \\
    &\hphantom{= \frac{1}{2} \norm{\hat{A}} \norm{\hat{B}} \sum_{\mathclap{V, V^\prime \in \{W, W^\dagger\}}} \big({}} + \imag \braket{\hat{V}_B^{\prime \, \dagger} \hat{V}_A} \sin \alpha_{\pm} \big) \,,
\end{align}
% Summation over $\hat{V}_A \in \{ \hat{W}_A, \hat{W}_A^\dagger \}$ and $\hat{V}_B \in \{ \hat{W}_B, \hat{W}_B^\dagger \}$, in combination with \cref{eq:qudit_dynamical_correlations},
where in the second line we have rearranged the terms in the sum.
Combining this expression with \cref{eq:probability_explicit} finally yields the desired result in \cref{eq:probability_corr}.%

\section{Variance of the Hadamard test}
\label{app:variance}
We proceed with the proof of \cref{eq:var_formula}. To this end, we rewrite \cref{eq:probability_corr} by expressing the probability $P_{V_A V_B}^{\pm} (t_1, t_2)$ of measuring $\ket{0}$ on the ancilla as the expectation value of the projector $\hat{\Pi}_0 = \ket{0}\bra{0} \otimes \mathds{1}$:
\begin{multline}
    C_{A B}^{\pm}(t_1,t_2) = \norm[\big]{\hat{A}} \norm[\big]{\hat{B}} \\
    \times\sum_{\mathclap{V, V^\prime \in \{W, W^\dagger\}}} \Big( \matrixelement[\big]{\Psi_{V_A V_B^\prime}^{\pm}(t_1, t_2)}{\hat{\Pi}_0}{\Psi_{V_A V_B^\prime}^{\pm}(t_1, t_2)} - \frac{1}{2} \Big) \,.
\end{multline}
Since the measurements of the four Hadamard-test configurations with controlled unitaries $\hat{V}_{A} \in \{ \hat{W}_A, \hat{W}^{\dagger}_A \}$ and $\hat{V}_{B} \in \{ \hat{W}_B, \hat{W}^{\dagger}_B \}$ are independent, standard propagation of variance yields~\cite{Polla2023}
% Therefore, the variance of the dynamical correlation function is
\begin{align}
   \variance[C^{\pm}_{A B}(t_1,t_2)] = \norm[\big]{\hat{A}}^2 \norm[\big]{\hat{B}}^2
    \sum_{\mathclap{V, V^\prime \in \{W, W^\dagger\}}}  \variance[\hat{\Pi}_0]^{\pm}_{V_A V_B^\prime}(t_1, t_2) \,.
\end{align}
If the Hadamard test with controlled unitaries $\hat{V}_A$ and $\hat{V}_B$ is repeated $M_{V_A V_B}$ times, the probability $P_{V_A V_B}^{\pm}(t_1, t_2)$ can be estimated with variance
\begin{multline}
    \variance[\hat{\Pi}_0]^{\pm}_{V_A V_B}(t_1, t_2)  \\
\begin{split}
    =\frac{1}{M_{V_A V_B}} \Big[ &\matrixelement[\big]{\Psi_{V_A V_B}^{\pm}(t_1, t_2)}{\hat{\Pi}_0^2}{\Psi_{V_A V_B}^{\pm}(t_1, t_2)} \\
    &- \matrixelement[\big]{\Psi_{V_A V_B}^{\pm}(t_1, t_2)}{\hat{\Pi}_0}{\Psi_{V_A V_B}^{\pm}(t_1, t_2)}^2 \Big] \,.
\end{split}
\end{multline}
For the projector $\hat{\Pi}_0$, the simple identity $\hat{\Pi}_0^2 = \hat{\Pi}_0$ holds, and thus for an arbitrary state
\begin{align}
    \braket{\hat{\Pi}_0^2} - \braket{\hat{\Pi}_0}^2 = \braket{\hat{\Pi}_0} \big( 1 - \braket{\hat{\Pi}_0} \big) \,.
\end{align}
Finally, identifying the expectation value $\langle \hat{\Pi}_0\rangle$ with the probability $P_{V_A V_B}^{\pm} (t_1, t_2)$ concludes the proof.

In order to arrive from \cref{eq:var_formula} at the \textit{a priori} upper bound 
$\variance[C^{\pm}_{A B}(t_1,t_2)] \leq M^{-1} \norm{\hat{A}}^2 \norm{\hat{B}}^2$, we used the fact that the function $f(x) = x(1-x)$ has a maximum value $f(x=1/2) = 1/4$ in the interval $[0,1]$.

\section{Decomposition for many-body observables}
\label{app:n-body_obs}

Typical $n$-spin observables are given by strings of the form
\begin{align}
    \hat{X} = \hat{S}^{\alpha_1}_1\otimes\cdots\otimes\hat{S}^{\alpha_n}_n \,,
\label{eq:X_prod_S}
\end{align}
where each $\hat{S}^{\alpha_k}_k$, with $\alpha_k \in \{x,y,z\}$, is a spin operator.
The unitary $\hat{W}_X$ can then be calculated using the unitary decomposition of each of the (one-body) spin operators:
\begin{align}
    \hat{S}^{\alpha_k}_k = \frac{1}{2}\norm{\hat{S}^{\alpha_k}_k}(\hat{W}_{\alpha_k} + \hat{W}^{\dagger}_{\alpha_k}) \,.
\label{eq:decomp_S}
\end{align}
Inserting Eq.~\eqref{eq:decomp_S} into Eq.~\eqref{eq:X_prod_S} and using the definition of $\hat{W}_X$, we get
\begin{align}
    \hat{W}_X = \hspace{1em} \sum_{\mathclap{\left\{ V_k = W_{\alpha_k}, W_{\alpha_k}^\dagger \right\}}} \hat{V}_1\otimes\cdots\otimes \hat{V}_n \,.
\end{align}
In this case, the experimental implementation of the controlled unitaries in the Hadamard-test simplifies since we have to deal with only a product of two-body operations.
However, the number of Hadamard tests to be performed increases exponentially with the length of the string in Eq.~\eqref{eq:X_prod_S}.

For a general multi-qudit observable, representations similar to strings of one-qudit observables can be employed to simplify the experimental implementation of controlled operations in the Hadamard test.
In practice, however, an approximate decomposition such as the one resulting from the LCU procedure~\cite{Chakraborty2024} may be more efficient in terms of circuit depth for implementations on NISQ devices.

\bibliography{ref}

\end{document}